\documentclass[aps,prl,reprint,superscriptaddress]{revtex4-1}

\usepackage[T1]{fontenc}
\usepackage[utf8]{inputenc}
\usepackage{gensymb}
\usepackage{graphicx}
\usepackage{braket}
\usepackage{amsmath,amsfonts,amssymb}
\begin{document}


\title{Non-Local Control of Single Surface Plasmon}



\author{Benjamin Vest}
\thanks{B. Vest and M.-C. Dheur contributed equally to this work.}

\author{Marie-Christine Dheur}
\thanks{B. Vest and M.-C. Dheur contributed equally to this work.}
\affiliation{Laboratoire Charles Fabry, Institut d'Optique, CNRS, Université Paris-Saclay, 91127 Palaiseau cedex, France.}

\author{\'Eloïse Devaux}
\author{Thomas W. Ebbesen}

\affiliation{Institut de Science et d'Ingénierie Supramoléculaire, CNRS, Université de Strasbourg, 67000 Strasbourg, France.}

\author{Alexandre Baron}

\affiliation{Centre de Recherche Paul Pascal, CNRS, 33600 Pessac, France.}

\author{Jean-Paul Hugonin}

\affiliation{Laboratoire Charles Fabry, Institut d'Optique, CNRS, Université Paris-Saclay, 91127 Palaiseau cedex, France.}

\author{Jean-Jacques Greffet}
\author{Gaétan Messin}

\author{François Marquier}
\email[e-mail : ]{francois.marquier@institutoptique.fr}
\affiliation{Laboratoire Charles Fabry, Institut d'Optique, CNRS, Université Paris-Saclay, 91127 Palaiseau cedex, France.}


\date{\today}

\begin{abstract}
Quantum entanglement is a stunning consequence of the superposition principle. This universal property of quantum systems has been intensively explored with photons, atoms, ions and electrons. Collective excitations such as surface plasmons exhibit quantum behaviors. For the first time, we report an experimental evidence of non-local control of single-plasmon interferences through entanglement of a single plasmon with a single photon. We achieved photon-plasmon entanglement by converting one photon of an entangled photon pair into a surface plasmon. The plasmon is tested onto a plasmonic platform in a Mach-Zehnder interferometer. A projective measurement on the polarization of the photon allows the non-local control of the interference state of the plasmon. Entanglement between particles of various natures paves the way to the design of hybrid systems in quantum information networks.
\end{abstract}


\maketitle


Surface plasmons polaritons (SPPs) are collective excitations of electrons with a mixed mechanical and electromagnetic character\cite{ritchie1968}
. Their quantum nature has been demonstrated by Powell and Swan \cite{powell1959} in the 1960’s. The fast development of nanophotonics and the need for compact integrated devices for quantum communication applications has revived the interest for SPP in the quantum regime. Testing the pioneer quantum optics experiment with plasmons has then become a motivation to challenge the limits of plasmonics in the quantum optics regime \cite{kolesov2009,akimov2007,cai2014,fakonas2014,fujii2014,heeres2013,dheur2016}. Among the fundamental quantum features, quantum entanglement \cite{einstein1935} has raised much interest  \cite{aspect1982,weihs1998,tittel1999,rowe2001,giustina2013,dimartino2014,hensen2015} as entangled pairs are a fundamental resource in quantum teleportation \cite{ekert1991}, in entanglement-based cryptography \cite{bennett1993} and in other protocols \cite{kimble2008}. The effect of plasmonic conversion has first been investigated by Altewischer et al. where the plasmon-assisted transmission of entangled photons was studied \cite{altewischer2002}. Since then, several groups have explored the plasmonic coherence properties and confirmed that polarization entanglement between two photons could be preserved when at least one of the energy quanta had had a plasmonic character onto a metallic device \cite{fasel2005,ren2006,fakonas2015}. Such experiments prove that quantum correlations should exist between photons before/after the plasmonic chip and a - mainly unknown - plasmonic state. 

	Here we report a first illustration of quantum non-locality involving a well-defined single SPP state entangled with a single photon state. More precisely, we entangle the two polarization modes of a photon with two spatial modes of a single SPP propagating along a metal-dielectric interface on a plasmonic chip. Therefore, measuring the polarization state of the photon affects the path followed by the single SPP on the chip. We probe directly the quantum state of this single SPP by recombining both paths on a plasmonic beam splitter, and observing the result of single SPP interferences at the output of the device. Thanks to quantum non-locality, the visibility of the interference fringes can be remotely controlled by performing projective measurements on the photon polarization state. To further check the role of the photon-SPP entanglement, we perform a control experiment using a non-entangled photon-plasmon pair.

\section{Source of entangled photon pairs}

\begin{figure*}

\includegraphics{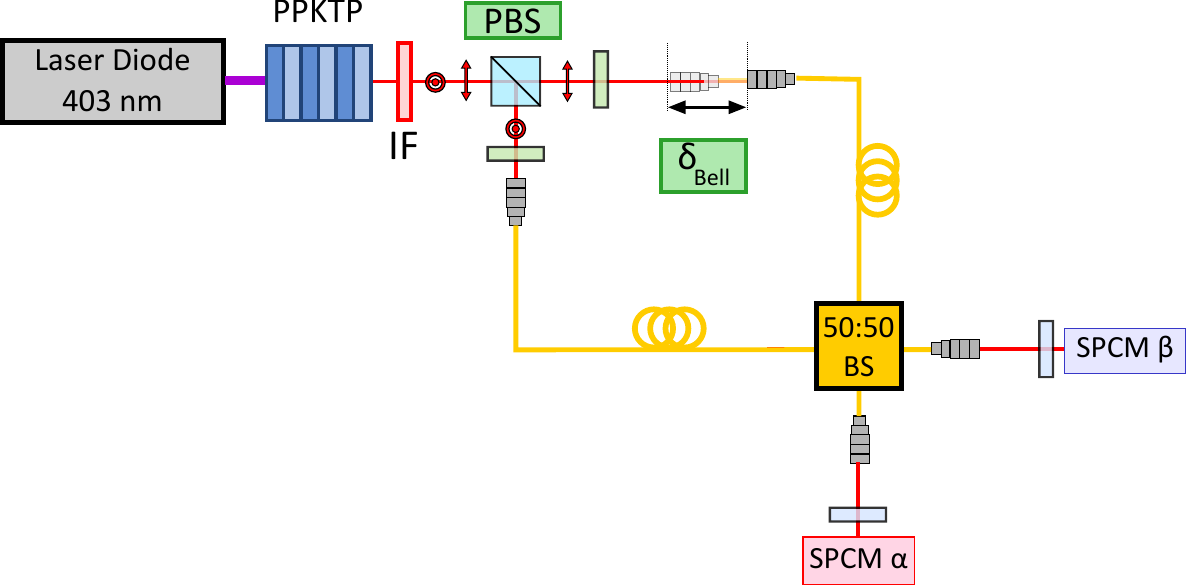}
\caption{\label{fig1}Sketch of the entangled photon source. The laser diode pumps a PPKTP crystal to generate pairs of red photons. They are separated by a PBS before being sent onto a 50:50 fibered beam splitter. The optical path difference between the photons is adjusted with a translation stage that mechanically adds a delay on one arm. We use the coincidences between SPCM $\alpha$ and SPCM $\beta$ to post-select the entangled state described in Eq. (2).}
\end{figure*}

The experimental setup is based on a source of entangled photon pairs, one of the photons being thereafter converted into a single plasmon. The source is a post-selected entangled photon pair source (PSEPPS). It consists in a spontaneous parametric down conversion (SPDC) source delivering 1 nm-spanned frequency-degenerate pairs of photons at 806 nm with linear orthogonal polarizations. The photons of the pair are indistinguishable except in polarization and are sent at each input of a 50:50 BS as in reference \cite{kuklewicz2004}. The photonic state can be then written considering the initial horizontal (H) or vertical (V) polarizations from the down-conversion process on the one hand, and the output modes $\alpha$ and $\beta$ of the fibered splitter on the other hand :

\begin{equation}
\ket{\psi_{\mathrm{out}}} = \frac{1}{2} (\ket{H_\alpha ; V_\alpha} +\ket{H_\alpha ; V_\beta} - \ket{H_\beta ; V_\alpha} - \ket{H_\beta ; V_\beta})
\end{equation}

By post selecting the coincidences between the output modes $\alpha$ and $\beta$ of the splitter, we reduce the state to an entangled state that can be cast in the form:

\begin{equation}
\ket{\psi_{\alpha \beta}} = \frac{\ket{H_\alpha ; V_\beta} - \ket{V_\alpha ; H_\beta}}{\sqrt{2}}
\end{equation}

Note that if the photons do not hit the beam splitter simultaneously, the arrival time of the photons on the last beam splitter contains information on the polarization so that the state is no longer an entangled state.

We can introduce a delay $\delta_{\mathrm{Bell}}$ between the two photons by mechanically translating one fiber's input along the photon path. This allows us to control the temporal indistinguishability of the photons. Optimal entanglement between photons is obtained for a delay that maximizes overlap of the photon wave packets when impinging on the beam splitter. Therefore, the degree of quantum entanglement can be progressively reduced by slightly moving the translation stage from this position.
To evaluate the quantum state produced by the PSEPPS, we estimated the violation of the Clauser, Horne, Shimony, and Holt (CHSH) form of Bell’s inequality by measuring the Bell parameter $S_{\mathrm{Bell}} $ of our source for different delays between the arrival times of the photons on the fibered beam splitter \cite{clauser1969}.We found the highest value $S_{\mathrm{Bell}} =2.44 \pm 0.04$ for the maximal temporal overlap of the photons. As $S_\mathrm{Bell}>2$, this value indicates that we clearly measure quantum correlations from a polarization entangled state. It is useful to note that such a state leads to strong correlations between photons in orthogonal polarizations : H and V. But more importantly, this property extends to correlations between any orthogonal linear polarizations in any other linear basis.

\section{Experimental setup}

\begin{figure*}

\includegraphics{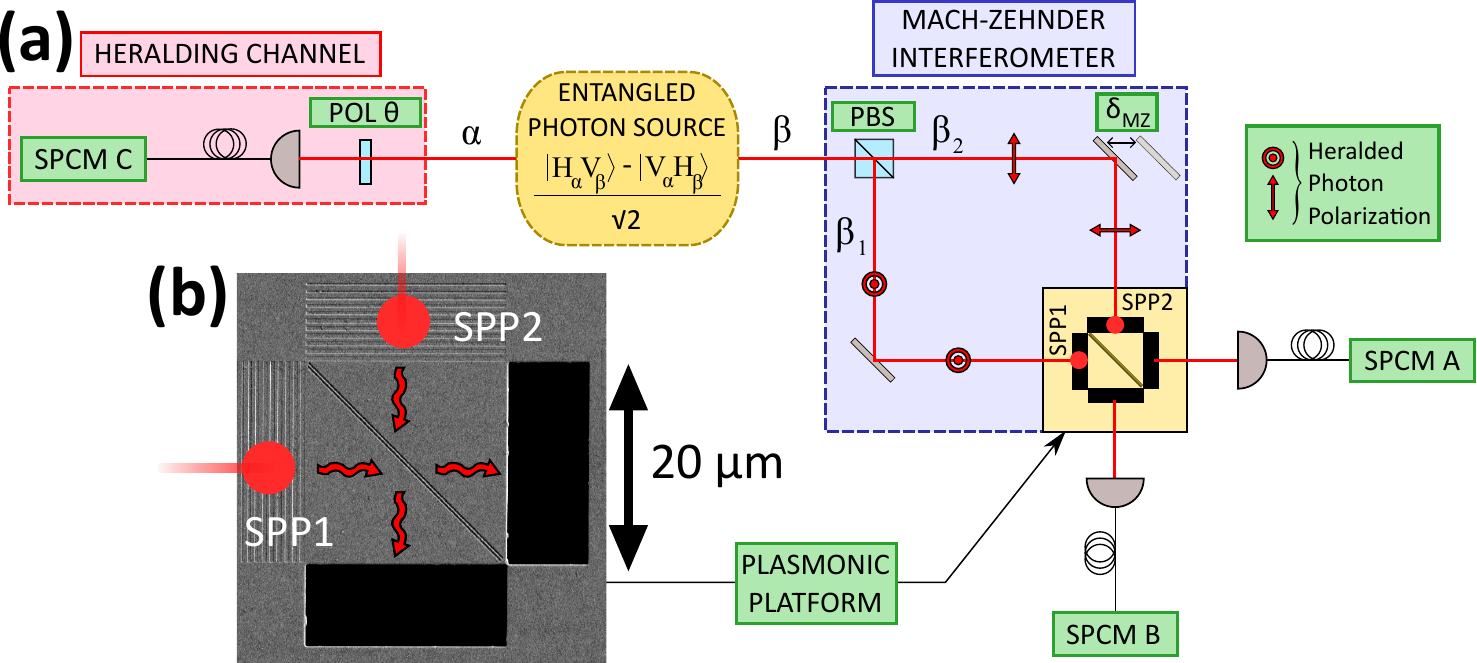}%
\caption{\label{fig2} a) Sketch of the experimental setup. A polarizer (POL) is placed on path $\alpha$ and a polarizing beam splitter (PBS) is placed on the path $\beta$. The PBS separates the incoming mode into two distinct photonic modes $\beta_1$ and $\beta_2$ which are next converted respectively into $\mathrm{SPP}_1$ and $\mathrm{SPP}_2$ modes on a plasmonic platform and recombined on a plasmonic beam splitter. This configuration is equivalent to a Mach-Zehnder (MZ) interferometer, with interference occuring on plasmonic modes. Fringes can be observed by introducing a variable delay on one of the photonic path of the interferometer. SPCM C detects heralding photon counts, SPCM A and SPCM B record detection counts heralded from SPCM C. b) SEM picture of the plasmonic platform (26). The top and left structures convert incident photonic modes into directional plasmonic modes. The surface plasmon beam splitter (SPBS) consists in two diagonal slits that can be seen in the center of the device's picture. The two black rectangles are out-coupling slits that convert SPPs back to photons. The latter are finally detected by SPCMs A and B.}
\end{figure*}

We will now describe the setup to manipulate the photon pairs and achieve photon-SPP entanglement. A simplified picture is shown on Figure \ref{fig1}, where each photon of the entangled pair is sent into opposite directions. The photon in mode $\alpha$ 
goes through a polarizer before being detected by the SPCM C and is used to herald the arrival of a photon in mode $\beta$ on SPCMs A and/or B. The photon in mode $\beta$ goes through a Mach-Zehnder (MZ) interferometer, which first separates polarization H and V using a polarizing beam splitter (PBS). The vertical (resp. horizontal) component of polarization is reflected (resp. transmitted), thus exciting the photonic mode $\beta_1$ (resp. $\beta_2$). We can now rewrite the state of the photon pair :

\begin{equation}
\ket{\psi_{\alpha \beta}} = \frac{\ket{H_\alpha ; 1_{\beta_1} ; 0_{\beta_2}} - \ket{V_\alpha ; 0_{\beta_1} ; 1_{\beta_2}}}{\sqrt{2}}
\end{equation}

Modes $\beta_1$ and $\beta_2$ are sent onto a plasmonic chip.  It is composed of five elements that are etched on a gold 300 nm-thick film on top of a silica substrate : two SPP directional launchers \cite{baron2011} that convert photonic modes $\beta_1$ and $\beta_2$ to plasmonic modes SPP1 and SPP2 (See Fig. \ref{fig2}(b)). The photon-to-SPP couplers send freely propagating SPPs onto a surface plasmon beam splitter (SPBS) made of two diagonal grooves. Two slits placed at the output of the SPBS convert the SPPs back into photons (black rectangles on Fig. \ref{fig2}(b)) which can be finally detected on SPCMs A and B.
The dimensions of all the components of the platform have been optimized with rigorous coupled wave analysis simulations \cite{silberstein2001}. The out-coupling slits dimensions are 10 $\mathrm{\mu m}$ in width and 20 $\mathrm{\mu m}$ long. The dimensions of the beam splitter (BS) have been optimized to produce similar amplitude and a $\frac{\pi}{2}$ phase shift between the reflection and transmission factors $r$ and $t$. The SPBS grooves are identical and their dimensions have been designed to be 180 nm-width and 140 nm-depth separated by a 140 nm gap. Measurement using a scanning electronic microscope (SEM) gives 171 nm-width and 145 nm-spacing. Due to the roughness of the gold at the bottom of the grooves, we can only provide an estimation of the depth of about 140 nm. We measured the SPBS intensity reflection and transmission factor $R = 17\%$ and $T=20\%$ respectively, leading to similar amplitudes $|r|\approx|t|$ . The losses $P=1-R-T$ are deduced to be 63\% and are mainly due to scattering processes. The phase difference between $r$ and $t$ has been measured using the difference between two fringes patterns recorded on SPCMs A and B. This phase difference is $100^{\circ} \pm 6^{\circ}$, which is close to the expected value $90^{\circ}$ although slightly different due to the lack of accuracy in the depth of the SPBS grooves. Just after the conversion of the photonic modes to plasmonic modes, the state of the photon-SPP  system before impinging on the SPBS can be finally written :

\begin{equation}
\ket{\psi'_{\alpha \beta}} = \frac{\ket{H_\alpha ; 1_{\mathrm{SPP}_1} ; 0_{\mathrm{SPP}_2}} - \ket{V_\alpha ; 0_{\mathrm{SPP}_1} ; 1_{\mathrm{SPP}_2}}}{\sqrt{2}}
\end{equation}

In other words, the state corresponds to a single photon entangled with a SPP. Note that the polarization degree of freedom of the photon in mode $\beta$ is converted into a path degree of freedom for the SPP. The photon in mode $\alpha$ is a heralding photon, and is sent into an \textit{ad hoc} channel. It will first fall onto a linear polarizer POL that projects the polarization of the photon before its detection by SPCM C. The photodetection on SPCM C opens a coincidence time window to detect photodetection events on SPCM A and B. The SPCM A and B are blind the rest of the time.
This projective measurement on $\alpha$ is used to post select the polarization of the particle in mode $\beta$  detected by SPCMs A and B. This polarization which is thus orthogonal to the polarization of the photon in mode $\alpha$.  Therefore, we excite photonic and plasmonic modes [$\beta_1-\mathrm{SPP}_1$] and [$\beta_2-\mathrm{SPP}_2$] with different amplitudes. A motorized translation stage introduces mechanically a delay $\delta_{\mathrm{HOM}}$ in one arm of the MZ interferometer.
 
 \section{Results}

In a first experiment we measure the heralding count rates on SPCMs A and B while rotating the heralding channel polarizer POL along different polarization directions $\theta$ . Writing the polarization state of the photon in mode $\alpha$ polarized along the direction $\theta$ in the $(H_{\alpha} ,V_{\alpha} )$ basis leads to:

\begin{equation}
\ket{\theta_\alpha} = \cos(\theta) \ket{H_\alpha} + \sin(\theta)\ket{V_\alpha}
\end{equation}

The projective measurement of the photon in mode $\alpha$ allows us to write the state of the particle in mode $\beta$ as:

\begin{align}
\ket{\psi'_\beta} &= \braket{\theta_\alpha | \psi'_{\alpha \beta}} \\ 
						   &=\cos(\theta) \ket{1_{\mathrm{SPP}_1} ; 0_{\mathrm{SPP}_2}} - \sin(\theta) \ket{0_{\mathrm{SPP}_1} ; 1_{\mathrm{SPP}_2}}
\end{align}

\begin{figure*}
\includegraphics{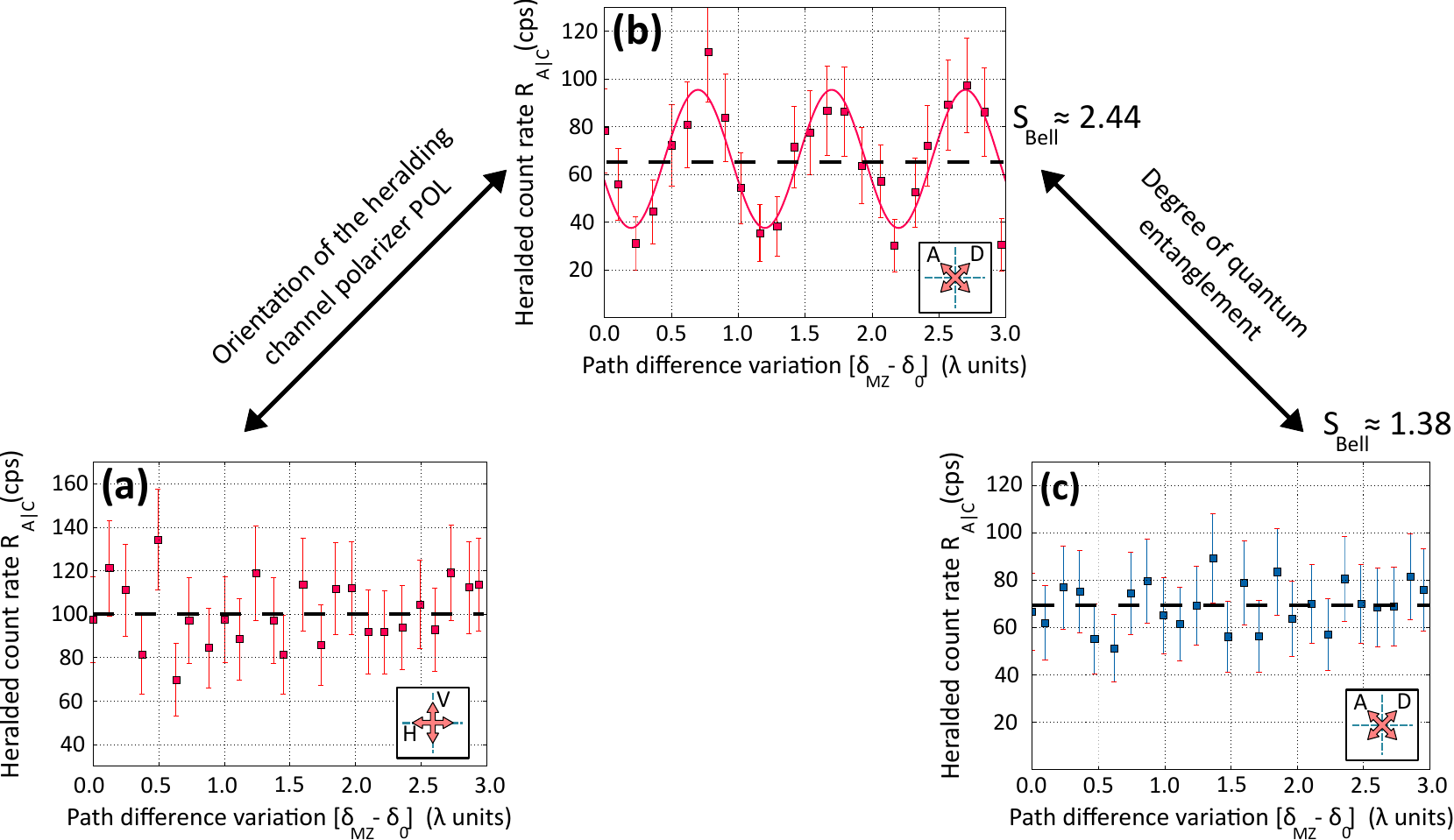}%
\caption{\label{fig3}(a) Heralded photon output rates $R_{A|C}$ as a function of the varying delay $\delta_{\mathrm{MZ}}$ between the MZ arms. (a) POL is aligned along H ($\theta=0^{\circ}$), one of the neutral axis of the PBS. No interference pattern appears for this direction of POL. (b) POL is aligned along the diagonal direction D ($\theta=45^{\circ}$, with respect to the neutral axis of the PBS). An interference pattern is clearly observed, showing that the single SPP excites indistinguishably the two input arms of the SPBS. The solid line is the sine fit function of the experimental data. For those two sets of data, the source has a Bell parameter around 2.44. (c), POL is still aligned in the diagonal/antidiagonal basis, but the source is now characterized by a Bell parameter $S_{\mathrm{Bell}} = 1.38$, meaning that the pairs are no longer entangled.}
\end{figure*}

We study how this projective measurement on photon in mode $\alpha$ affects the correlations with the output signal of the interferometer. Figure \ref{fig3} highlights two particular configurations. First, the direction $\theta$ of the heralding channel polarizer is chosen along one of the neutral axis of the input PBS (denoted as H or V on Fig. \ref{fig3}(a)). Let us choose direction V (resp . H). Thus, the photon in mode $\beta$ will be measured in a polarization orthogonal to state V (resp. H), meaning H (resp. V) : the photon will be for instance only transmitted by the PBS and will follow the path $\beta_2$ with 100\% probability : from the detection point of view, the \textit{which path} information is completely known. At the output of the SPBS, interferograms do not exhibit any visible fringes (see Fig \ref{fig3}(a)). We measure a stable heralding counting rate with respect to the path difference $\delta_{\mathrm{MZ}}$. The fluctuations of the heralded counts that are observed are mainly attributed to the photon detection noise.

We now rotate the polarizer POL by $45^{\circ}$ and align it along a diagonal direction D, so the test photon $\beta$ is now measured along the antidiagonal direction A, at $45^{\circ}$ with respect to the neutral axis of the input PBS. In this configuration, the photon has a 50\% probability to be either reflected or transmitted by the PBS, and both modes of the interferometer are excited with similar amplitudes : the PBS behaves like a 50:50 BS. Recombining the modes on the balanced SPBS allows to erase the which path information. Heralded counts interferograms on SPCM A and SPCM B clearly exhibits fringes that emerge from the signal background (see Fig \ref{fig3}(b)). Hence, the outcome of the plasmonic interferometer in terms of fringes visibility depends on the measurement performed on the photon $\alpha$ : this experiment displays a non-local control of the surface-plasmon interference state. We stress the fact that the state of the mode $\beta$ is chosen by the detection event by SPCMs A and B. As no relative time dependance between the particles is necessary, Eq.(5) suggests that the SPP state could be controlled in a delayed-choice experiment using the orientation of POL. However the visibility of the fringes is limited to $50 \pm 7 \%$. We explain this value by several origins. The first source of degradation comes from the plasmonic platform itself. The non-guided plasmonic modes on the film do not offer a perfect control of SPP modes overlap on the SPBS. With a classical source, the best alignment of the interferometer gave us no more than 80\% for the visibility of the SPP fringes. The second source of degradation comes from the quality of the entangled photon source itself. A perfect entangled source would have given $S_{\mathrm{Bell}}=2\sqrt{2}$. In our case we mainly attribute the origin of the measured value  ($S_{\mathrm{Bell}}=2.44 \pm 0.04$) to the poor mode overlap of the wave packets in the nonlinear crystal. Finally, the degradation of the visibility is also affected by the mechanical instability of our interferometer that smear the data in time. In our setup, detected SPPs can not experience pure dephasing processes, as they are much more likely to be absorbed first. Because we perform post-selection on surviving particles, quantum decoherence effects are negligible.

The previous results show a strong correlation between the choice of $\theta$ and the appearance of plasmonic fringes. The projective measurement on particle $\alpha$ allows a non-local control of the outcome of the plasmonic interferometer which is an indication of the entanglement between the photon and the plasmon.

As a further confirmation of the role of entanglement in the observation of interferences, we repeated the interference experiment for a lower degree of quantum entanglement. The polarizer POL is still aligned in the diagonal basis. We adjust the entangled photon source in order to operate with a Bell parameter as low as $1.38 \pm 0.04$ instead of $2.44 \pm 0.04$. For those settings, the state is no longer entangled. We report in Fig. \ref{fig3}(c) the vanishing of interferences in the diagonal/antidiagonal basis.

\section{Conclusion}
In summary, we have reported the generation of a hybrid entangled state consisting in a photon and a surface plasmon. This state is generated by exciting a plasmon with one of the photons of an entangled photon pair. Measurements performed on the state of polarization of the photon results in a non-local control of plasmonic interferences. We have observed this non local behavior by measuring the output of a plasmonic Mach-Zehnder interferometer. We have further checked that when replacing the polarization entangled photon pair by a mixed crossed-polarization photon pair, the non-local control is suppressed. These results pave the way to the development of hybrid plasmon-photon systems for quantum protocols. 

\subsection{Methods}

\textbf{Detection method.} AAll the photons in these experiments were sent to SPCMs, which deliver transistor-transistor logic pulses. SPCMs A and B are Perkin-Elmer modules (SPCM AQRH-14), and SPCM C is a Laser Component SPCM (Count-100C FC). To count the correlations between the heralding signal and the SPCMs A and B pulses, we used a PXI Express system from National Instruments (NI). The NI system is composed of a PXIe-1073 chassis on which NI FlexRIO materials are plugged : a field programmable gate array (FPGA) chip (NI PXIe-7961R) and an adapter module at 100 MHz (NI 6581). The FPGA technology allows changing the setting of the acquisition by simply programming the FPGA chip to whatever set of experiments we want to conduct. A rising edge from SPCM C triggers the detection of another rising edge on channel A or B or both at specific delays. Counting rates and correlations of heralded coincidences between channels A and B are registered. The resolution of the detection system is mainly ruled by the acquisition board frequency clock at 100 MHz, which corresponds to a time resolution of 10 ns.

\textbf{Photon pair source.} A potassium titanyl phosphate crystal (PPKTP crystal from Raicol) crystal is pumped at 403 nm by a tunable laser diode (Toptica). It delivers a 38mW powered-beam, focused in the crystal by a 300 mm focal length planoconvex lens. The waist in the crystal is estimated to be 60 $\mathrm{\mu}$ m. The crystal generates pairs of orthogonally polarized photons at 806 nm. The waist in the crystal is conjugated to infinity with a 100-mm focal-length plano-convex lens, and the red photons emerging from the crystal are separated in polarization by a polarizing beam splitter (PBS) cube (Fichou Optics). We remove the remaining pumping signal with an interferometer filter (IF) from AHF (FF01-810/10). The 806 nm photons are then coupled to a 50:50 fiber optic coupler. Each photon can be either transmitted or reflected with equal probabilities. 

\textbf{Mach-Zehnder interferometer} After these characterizations steps, the output mode $\beta$ of the entangled photon source is connected to the MZ interferometer. The photons in mode $\beta$ are coupled to polarization-maintaining monomode fibers (P1-780PM-FC) via collimators (F220FC-780, Thorlabs). Each photon is outcoupled via Long Working Distance M Plan Semi-Apochromat microscope objectives (LMPLFLN-20X BD, Olympus) and sent to two different inputs of a PBS (Fichou Optics) with orthogonal polarizations. They leave the cube by the same output port and were focused with a 10X microscope objective (Olympus) on the plasmonic sample. The plasmonic sample is mounted on a solid immersion lens. The surface plasmons propagating on the chip leave the sample by two orthogonal output slits. The conversion of the SPPs back to photons via the slits leads to two different directions of propagation in free space. The photons from the output ports are collected from the rear side of the sample using mirrors and a 75-mm focal-length lens for each output. The output modes are then conjugated to multimode fibers via a 10X microscope objective (Olympus), which are connected to the SPCMs.

\textbf{Plasmonic platform sample fabrication.} We deposited 300-nm-thick gold films on clean glass substrates by e-beam evaporation (ME300 Plassys system) at a pressure of $2.10^{-6}$ mbar and at a rate of 0.5 nm/s. The rms roughness is 1 nm. The films were then loaded in a crossbeam Zeiss Auriga system and milled by a focused ion beam at low current (20 pA), except for the large slits used to decouple plasmons for propagating light that were milled at 600 pA.

\subsection{Acknowledgments}

We would like to show our gratitude to Philippe Lalanne and Jean-Claude Rodier for their crucial role in this project. We also acknowledge E. Rousseau, F. Cadiz, and N. Schilder for their help in the beginning of this study, as well as L. Jacubowiez, A. Browaeys, and P. Grangier for fruitful discussions. 

The research was supported by a DGA-MRIS (Direction Générale de l’Armement– Mission Recherche et Innovation Scientifique) scholarship, by RTRA (Réseau Thématique de Recherche Avancée) Triangle de la Physique, by the SAFRAN-IOGS chair on Ultimate Photonics and by a public grant overseen by the French National Research Agency (ANR) as part of the « Investissements d’Avenir » program (reference: ANR-10-LABX-0035, Labex NanoSaclay). J.-J.G. acknowledges the support of Institut Universitaire de France. 

\subsection{Author contributions}

G.M., F.M., and J.-J.G. initiated the project. J.-P.H. designed the chip, which was fabricated by E.D. and T.W.E. and characterized by A.B. and M.-C.D. M.-C.D. built the setup. Quantum experiments and data analysis were performed by B.V. and M.-C. D. under the supervision of F.M., G.M. and J.-J. G.. B.V. , M.-C.D., F.M., and J.-J.G. wrote the paper and discussed the results.

\subsection{Competing financial interests}

The authors declare no competing financial interests.


%

\end{document}